\documentclass[a4paper,12pt]{article}
\usepackage{graphicx} % standard LaTeX graphics tool
\usepackage{amsfonts,amsmath,amsthm,color}
\newcommand{\R}{\mathbb{R}}
\newcommand{\C}{\mathbb{C}}
\newcommand{\N}{\mathbb{N}}

\newtheorem{defin}{Definition}
\newtheorem{teo}{Theorem}
\newtheorem{prop}{Proposition}
\newtheorem{rem}{Remark}
\newtheorem{coro}{Corollary}

\newtheorem{lem}{Lemma}
\newtheorem{example}{Example}
\begin{document}
\title{Neumark Operators and Sharp Reconstructions, the finite dimensional case\footnote{Copyright (2007) American Institute of Physics. This article may be downloaded for personal use only. Any other use requires prior permission of the author and the American Institute of Physics.
The following article appeared in Journal of Mathematical Physics, 48, 022102 (20007) and may be found at http://link.aip.org/link/?jmp/48/022102 }}
\author{Roberto Beneduci\footnote{Elettronic-mail rbeneduci@unical.it}\\
{\em Dipartimento di Matematica, Universit\`a della Calabria,}\\
{\em Istituto Nazionale di Fisica Nucleare, Gruppo c. Cosenza,}\\
{\em Arcavacata di Rende (Cs), Italy}\\
}
\date{}
\maketitle
\begin{abstract}
\noindent
A commutative POV measure $F$ with real spectrum is characterized by the existence of a PV measure $E$ (the sharp reconstruction of $F$) with real spectrum such that $F$ can be interpreted as a randomization of $E$. This paper focuses on the relationships between this characterization of commutative POV measures and Neumark's extension theorem. In particular, we show that in the finite dimensional case there exists a relation between the Neumark operator corresponding to the extension of $F$ and the sharp reconstruction of $F$. The relevance of this result to the theory of non-ideal quantum measurement and to the definition of unsharpness is analyzed.
\end{abstract}
\pagestyle{plain}
\vskip3cm
%%%%%%%%%%%%%%%%%%%%%%%%%%%%%%%%%%%%%%%%%%%
\renewcommand{\thesection}{\Roman{section}.}
\section{Introduction}
A Positive Operator Valued measure (POV measure) 
is a map $F$ %:{\cal B}(\R)\to{\cal F(H)}$, 
from the Borel $\sigma$-algebra of the reals ${\cal B}(\R)$ into the set of bounded, positive, self-adjoint operators ${\cal F(H)}$ in a Hilbert space $\cal H$.
POV measures were introduced in quantum mechanics by E. B. Davis$^1$, G. Ludwig$^2$ and A. S. Holevo$^{3-5}$ in order to generalize the concept of observable of a physical system. Before the introduction of POV measures, the observables were described by Projection Valued measures (PV measures). In the new terminology, we distinguish between sharp observables, which are described by PV measures, and unsharp observables, which are described by POV measures.

\noindent
Although the introduction of POV measures comes from the foundational analysis of quantum mechanics, POV measures find several applications, for example in quantum stochastic processes, quantum optics$^{6,7}$ and relativistic quantum mechanics$^8$. A way to justify their use is to consider the process of repeated measurements of a quantum observable$^9$ or to derive them as a consequence of the probabilistic structure of Quantum Mechanics$^{3,4,5,10}$. 

\noindent
The state of a physical system can be represented by a density operator $\rho$ acting in $\mathcal{H}$.
A. S. Holevo$^{3-5}$ has shown that there exists a one-to-one correspondence between POV measures
and affine maps from the set of the states of a physical system into the set of probability measures on ${\cal B}(\R)$.
The affine map $\rho \mapsto\mu_{(\cdot)}^F(\rho)$ corresponding to the POV measure $F$ is determined by the relation
\begin{equation}
\mu_{(\Delta)}^F(\rho)=Tr[\rho F(\Delta)], \quad \text{for all }\Delta\in\mathcal{B}(\R).
\end{equation}

\noindent
This allows one to interpret the real number $\mu_{(\Delta)}^F(\rho)=Tr[\rho F(\Delta)]$
as the probability that the outcomes of the measurement of an unsharp observable$^{1,2,3,9,10}$ $F$ is in $\Delta$ when the physical system under consideration is in the state $\rho$.
Equation 1 generalizes the analogous equation 
$$\mu_{(\Delta)}^E(\rho)=Tr[\rho E(\Delta)]$$
which has the same meaning for a sharp observable$^{1,2,3,9,10}$ $E:{\cal B}(\R)\to\cal E(H)$,
from ${\cal B}(\R)$ to the set of projection operators $\cal E(H)$. Therefore, the unsharp observables, represented by POV measures, generalize the sharp ones, represented by PV measures. 

Several characterizations of POV measures$^{11-17}$ can be found in the literature.

This paper focuses on the problem of finding a relation between two of them. The first characterization is due to M. A. Neumark$^{11}$ and establishes that every POV measure $F:{\cal B}(\R)\to{\cal F(H)}$ can be extended to a PV measure $E^+:{\cal B}(\R)\to{\cal F}({\cal H}^+)$ in an extended Hilbert space ${\cal H}^+$, in such a way that $F$ is the projection of $E^+$ on $\cal H$ (see theorem 3). Neumark's theorem brings to a physical
interpretation$^{3,4}$ of the measurement of non-orthogonal POV measures (see corollary 1).

\noindent
The second characterization can be found in Ref.s $15-18$, and concerns commutative POV measures.
In Ref. 17 (see theorem 2), it is shown that for each commutative POV measure $F$ there exists a self-adjoint operator $A$ with spectrum  $\sigma(A)$, named the sharp reconstruction of $F$, such that, for every $\lambda\in\sigma(A)$, there is a probability measure $\mu^A_{(\cdot)}(\lambda):{\cal B}(\R)\to[0,1]$, such that, 
\begin{equation}
F(\Delta)=\int\mu^A_{(\Delta)}(\lambda)dE^A_{\lambda}=\mu^A_{\Delta}(A),\quad\Delta\in{\cal B}(\R),
\end{equation}
where $E^A$ is the PV measure corresponding to $A$ and the rules of the functional calculus has been used (see pages 7 and 8 in section II and Ref. 19). The sharp reconstruction $A$ is unique up to bijections. This characterization allows$^{16,17}$ one to interpret a commutative POV measure $F$ as a randomization of its sharp reconstruction $A$ (see theorem 2 and comments to the theorem).

In the present paper, we restrict ourself, without loss of generality, to the case of POV measures with spectrum in $[0,1]$ (see the introduction to section III and appendix A). Let $F$ be a commutative POV measure, $A=\int\lambda dE^A_{\lambda}$ its sharp reconstruction, $E^+$ a Neumark extension of $F$ and $A^+=\int\lambda dE^+_{\lambda}$ the  corresponding operator. We show (theorem 4) that for any bounded, measurable function $f:[0,1]\to\R$ there exists a bounded, measurable function $G_f:[0,1]\to[0,1]$ such that $$P^+f(A^+)_{\vert{\cal H}}=G_f(A),$$
where $P^+$ is the operator of projection onto $\cal H$, $G_f(A)=\int G_f(\lambda)dE^A_{\lambda}$, $f(A^+)=\int f(\lambda)dE^+_{\lambda}$ and $f(A^+)_{\vert{\cal H}}$ is the restriction of $f(A^+)$ to $\cal H$.  

Moreover, we prove (theorem 7) that, in the finite dimensional case, there exists a bounded, one-to-one function $f$ such that $G_f$ is one-to-one (this is shown to be true for POV measures both with finite and countably infinite outcome sets). This gives a notion of equivalence between sharp reconstructions and projections of Neumark operators which generalizes the one proposed in Ref. 18. We denote this equivalence by $A\leftrightarrow \Pr A^+$ (see definition 9). This result suggests that it is reasonable to look for an extension of theorem 7 to the infinite dimensional case.  
Furthermore, it bears interesting implications to the theory of non-ideal quantum measurement$^{20,21}$ (see corollary 3).

Finally, the properties of the sharp reconstruction proved in theorems 6 and 7 allow us to comment on the differences between the definition of unsharpness proposed in Ref.s 20,21 (see definition 11) and that given in Ref.s 22-25 (see definition 12). In particular, we show that the two definitions do not coincide even in the case of commutative POV measures. Moreover, denoting by ${\cal A}_1$ and ${\cal A}_2$ the sets of sharp observables of which $F$ is an unsharp version according to definitions 11 and 12 respectively, it is possible to see that ${\cal A}_1\cap{\cal A}_2\neq\emptyset$. In fact, the sharp reconstruction belongs to ${\cal A}_1\cap{\cal A}_2$. Furthermore, we show that definition 12 can be modified in order to enlarge ${\cal A}_2$ to a set which contains ${\cal A}_1\cup{\cal A}_2$ (see definition 13 and theorem 9).

The paper is organized as follows. In section II we introduce some basic definitions, state the classical theorem of Neumark and summarize the main results of Ref.s 16, 17. In subsection III.1 we state theorem 4 and present some examples of POV measures such that $A\leftrightarrow \Pr A^+$.
In subsection III.2, we prove the main result of the paper, theorem 7, and give some examples. Then, in the last section, we prove corollary 3 and make some observations on the definition of unsharp observable. In the appendix A we show that we can restrict ourself, without loss of generality, to POV measures with a bounded spectrum. In the appendix B we prove theorem 4. In the appendix C we prove lemma 1. 
\section{Preliminaries}
In this section we fix the basic notation and terminology, state the classical theorem of Neumark and give the characterization of commutative POV measures obtained in Ref.s 16-17.  

We denote by ${\cal B}(\R)$ the Borel $\sigma$-algebra of $\R$, by $\textbf{0}$ and $\textbf{1}$ the null and the identity operators respectively, by ${\cal L}_s({\cal H})$ the space of all bounded self-adjoint linear operators acting on a Hilbert space $\cal H$ with scalar product $\langle\cdot,\cdot\rangle$, by ${\cal F}({\cal H})\subset{\cal L}_s({\cal H})$ the subspace of all positive, bounded self-adjoint operators on $\cal H$, by ${\cal E}({\cal H})\subset{\cal F}({\cal H})$ the subspace of all projection operators on $\cal H$.     

\noindent
\begin{defin}
A Positive
Operator Valued measure (for short, POV measure) is a map $F:{\cal B}(\R)\to{\cal
F}({\cal H})$ such that:
\begin{enumerate}
    \item[]  if $\{\Delta_n\} \,\,\hbox{is a countable family of disjoint
    sets in}\,\, {\cal B}(\R)$ then
    \begin{equation}
    F\big(\bigcup_{n=1}^{\infty}\Delta_n\big)=\sum_{n=1}^{\infty}F(\Delta_n),
    \end{equation}
    \hbox{where the series converges in the weak operator topology}.
    \end{enumerate}
\end{defin}
\begin{defin}
    A POV measure is said to be: 
\begin{itemize}
    \item[{1.}] normalized if  
    \begin{equation}
    F(\R)={\bf{1}}.
    \end{equation}
    \item[{2.}] commutative if
    \begin{equation}
    \big[F(\Delta_1),F(\Delta_2)\big]={\bf{0}},\,\,\,\,\forall\,\Delta_1\,,\Delta_2\in{\cal B}(\R). 
    \end{equation}
    \item[{3.}]  orthogonal if
    \begin{equation}
    F(\Delta_1)F(\Delta_2)={\bf{0}}\,\,\,\hbox{if}\,\,\Delta_1\cap\Delta_2=
    \emptyset. 
    \end{equation}
\end{itemize}
\end{defin}
\noindent
In what follows we shall always refer to normalized POV
measures defined on ${\cal B}(\R)$.
\begin{defin}
A Projection Valued measure (for short, PV measure) is an orthogonal, normalized POV measure.
\end{defin}
\noindent
It is simple to see that for a PV measure $E$, we have $E(\Delta)=E(\Delta)^2$, for any $\Delta \in \mathcal{B}(\R)$. Then, $E(\Delta)$ is a projection operator for every $\Delta\in{\cal B}(\R)$, and the PV measure is a map $E:{\cal B}(\R)\to{\cal E}({\cal H})$. 

\noindent
In quantum mechanics, non-orthogonal normalized POV measures are also called \textbf{generalised} or \textbf{unsharp} observables and PV measures \textbf{standard} or \textbf{sharp} observables. 

\noindent
We shall use the term ``measurable'' for the Borel measurable functions.
For any vector $x\in\cal H$ the map 
$$\langle F(\cdot)x,x\rangle \,:\,\mathcal{B}(\R)\to \R ,
\qquad
\Delta \mapsto \langle F(\Delta)x,x\rangle,$$
is a Lebesgue-Stieltjes measure. There exists a one-to-one correspondence$^{26}$ between POV measures $F$ and POV functions $F_{\lambda}:=F((-\infty,\lambda])$. In the following we will use the symbol $d\langle F_{\lambda}x,x\rangle$ to mean integration with respect to the measure $\langle F(\cdot)x,x\rangle$. 

\noindent
We shall say that a function $f:\R\to\R$ is bounded with respect to a POV measure $F$, if it is equal to a bounded function $g$ almost everywhere (a.e.) with respect to $F$, that is, if $f=g$ a.e. with respect to the measure $\langle F(\cdot)x,x\rangle$,  $\forall x \in \mathcal{H}$. % runs through all the vectors of the space $\cal H$. 
For any real, bounded and measurable function $f$ and for any $F\in{\cal F}({\cal H})$, there is a unique$^{27}$ bounded self-adjoint operator $B\in{\cal L}_s({\cal H})$ such that
\begin{equation}
\langle Bx,x\rangle=\int f(\lambda)d\langle F_{\lambda}x,x\rangle,\quad\text{for each}\quad x\in\cal H.
\end{equation} 
If equation (7) is satisfied, we write $B=\int f(\lambda)dF_{\lambda}$. 
\noindent
\begin{defin}
The spectrum $\sigma(F)$ of a POV measure $F$ is the closed set
$$\left\{\lambda\in\R:\,F\big((\lambda-\delta,\lambda+\delta)\big)\neq
0,\,\forall\delta>0,\,\,\right\}.$$
\end{defin}
By the spectral theorem$^{19,28}$, PV measures are in a one-to-one correspondence with self-adjoint operators. 
In fact, we recall the following theorem of functional analysis.
\begin{teo}[see Ref.s $19$]
There is a one-to-one correspondence between self-adjoint operators $B$ on a Hilbert space $\cal H$ and PV measures $E^B$ on $\cal H$, the correspondence being given by
$$B=\int\lambda dE^B_{\lambda}.$$
\end{teo}
\noindent  
In the following we do not distinguish between PV measures and the corresponding self-adjoint operators. 

If $f:\R\to\R$ is a measurable real-valued function, then we can define the self-adjoint operator$^{19}$
$$f(B)=\int f(\lambda) dE^B_{\lambda}.$$
If $f$ is bounded, then $f(B)$ is bounded.$^{19}$ 
\begin{defin}
Two bounded self-adjoint operators $A$ and $B$ are said to be equivalent if there exists a bounded, one-to-one, measurable function $f$ such that $A=f(B)$. In this case we write $A\leftrightarrow B$.
\end{defin}

\begin{defin} We say that the triplet $\big(F,B,\mu^B_{(\cdot)}(\lambda)\big)$ satisfies the thesis of von Neumann theorem$^{29-31}$ if $\mu_{(\Delta)}(B)=F(\Delta)$, for every $\Delta\in{\cal B}(\R)$. 
\end{defin}

\noindent
Summing up the results obtained in Ref.s 15-17 we can state the following theorem:
\begin{teo}[see Ref. 16,17]
A POV measure $F:{\cal B}(\R)\to{\cal
F(H)}$ is commutative if and only if there exist a self-adjoint operator $B$ and, for every $\lambda\in\sigma(B)$, a probability measure$^{32,33}$ $\mu^B_{(\cdot)}(\lambda):{\cal B}(\R)\to[0,1]$ such that the triplet $\big(F,B,\mu^B_{(\cdot)}(\lambda)\big)$ satisfies the thesis of von Neumann's theorem.

\noindent
Moreover, there exists a couple $\big(A,\mu^A_{(\cdot)}(\lambda)\big)$ such that: i) the triplet $\big(F,A,$ $\mu^A_{(\cdot)}(\lambda)\big)$ satisfies the thesis of von Neumann's theorem; ii) for every triplet $\big(F,B,\mu^B_{(\cdot)}(\lambda)\big)$ satisfying the thesis of von Neumann's theorem, there exists a real function $g$ such that $A=g(B)$. The operator $A$ is unique up to bijections.
\end{teo} 
\begin{defin} 
The operator $A$ defined by Theorem 2 (or, equivalently, the corresponding PV measure $E^A$) is called the sharp reconstruction of $F$. 
\end{defin} 
\noindent
Theorem 2 suggests to interpret$^{16,17}$ the outcomes of the measurement of $F$ as deriving from a randomization of the outcomes of the measurement of its sharp reconstruction $E^A$. Indeed, for every $\Delta\in{\cal B}(\R)$ and $\lambda\in\sigma(A)$, $\mu^A_{(\Delta)}(\lambda)$ can be interpreted as the probability that the outcome of a measurement of $F$ is in $\Delta$ when the outcome of the measurement of $E^A$ is $\lambda$. 

\begin{teo}[Neumark$^{29,31}$]
Let $F$ be a POV measure of the Hilbert space $\cal H$. Then, there exist a Hilbert space $\cal H^+\supset\cal H$ and a PV measure $E^+$ of the space $\cal H^+$ such that
$$F(\Delta)=P^+E^+(\Delta)_{\vert{\cal H}}$$
where $P^+$ is the operator of projection onto $\cal H$.
\end{teo}
\begin{defin}
Each operator $\int f(\lambda)dE^+_{\lambda}$, where $f$ is a one-to-one, measurable, real valued function, is said to be a Naimark operator corresponding to $F$. The Neumark operator $\int\lambda dE^+_{\lambda}$ is denoted by $A^+$.
\end{defin}
\noindent
The following corollary yields a physical
interpretation of the measurement of a non-orthogonal POV measure.
\begin{coro}[see Ref.s 3,4]

For any POV measure $F:{\cal B}(R)\to{\cal F}({\cal H})$ there exist a Hilbert space ${\cal H}_0$, a pure state $S_0$ in ${\cal H}_0$ and a PV measure $E^+:{\cal B}(R)\to{\cal E}({\cal H}^+)$ in ${\cal H}^+={\cal H}\otimes{\cal H}_0$
 such that
\begin{equation}
\mu_{(\Delta)}^{E^+}(S\otimes S_0)=\mu_{(\Delta)}^F(S)\quad \Delta\in{\cal B}(R)
\end{equation}
for each state $S$ in $\cal H$. The converse is also true, that
is, for every triplet (${\cal H}_0$, $S_0$, $E^+$), where $E^+$ is
a PV measure in the Hilbert space ${\cal H}\otimes{\cal H}_0$
and $S_0$ is a pure state in ${\cal H}_0$, there exists a unique
POV measure $F$ satisfying equation (8).
\end{coro}

\noindent
Equation (8) establishes the existence of a pure state $S_0$ such  
that the measurements of the observables $F$ and $E^+$ are
statistically equivalent.
Therefore the measurement of an unsharp observable in the Hilbert space $\cal H$ is equivalent
to the measurement of a sharp one in the Hilbert space ${\cal H}\otimes{\cal H}_0$ which
represents the composition between the system and additional independent degrees
of freedom described by ${\cal H}_0$. 

\begin{prop} 

Let us consider the extension $E^+$ of a POV measure $F$ and the Neumark operator $A^+=\int\lambda dE^+_{\lambda}$ corresponding to $E^+$. Let $f$ be a measurable function which is bounded with respect to $E^+$. Then 
$$P^+f(A^+)_{\vert\cal H}=\int f(\lambda)dF_{\lambda}$$
and $P^+f(A^+)_{\vert\cal H}$ is a bounded self-adjoint operator.
\end{prop} 
\begin{proof}
\begin{equation*}
\langle P^+f(A^+)x,y\rangle=\int_{-\infty}^{\infty}f(\lambda)d\langle E^+_{\lambda}x,P^+y\rangle=\int_{-\infty}^{\infty}f(\lambda)d\langle F_{\lambda}x,y\rangle
\end{equation*}
for every $x,y\in\cal H$.

\noindent
The boundedness and the self-adjointness of $P^+f(A^+)_{\vert\cal H}$ come, respectively, from the boundedness and the real-valuedness of $f$ with respect to $E^+$ (see theorem 10 in Ref. 27).   
\end{proof}
\begin{defin}

Whenever there exists a one-to-one, bounded, measurable function $f:\sigma(A^+)\to\R$ such that the sharp reconstruction $A$ of a commutative POV measure $F$ is equivalent to $P^+f(A^+)_{\vert\cal H}$ we write $A\leftrightarrow \Pr A^+$ and say that the sharp reconstruction $A$ is equivalent to the projection of a Neumark operator corresponding to $F$. 
\end{defin}

\noindent
\section{Naimark Operators and Sharp Reconstructions}
In the present section we analyze the relationships between Neumark's theorem and theorem 2. 
In subsection III.1, generalizing a result obtained in Ref. 18, we show that for every commutative POV measure $F$ and for every bounded and measurable function $f$ there exists a function $G_f$ such that $G_f(A)=P^+f(A^+)_{\vert\cal H}$, where $A$ and $A^+$ are respectively the sharp reconstruction of $F$ and the Neumark operator $\int \lambda dE^+_{\lambda}$ corresponding to the extension $E^+$ of $F$. We also give some examples where $A\leftrightarrow Pr A^+$. In subsection III.2 we prove that, in the finite dimensional case, a positive answer can be given to the problem of the equivalence between the sharp reconstruction $A$ and the projection of the Neumark operator, i.e., there exists always a one-to-one, bounded and measurable function $f$ such that $A\leftrightarrow P^+f(A^+)_{\vert\cal H}$.

\noindent
We recall that, as shown in Ref. 15, the sharp reconstruction $A$ of a given POV measure $F$ coincides with the sharp reconstruction of the POV measure $\bar{F}$ defined by $\bar{F}(\Delta)=F\big(f(\Delta\cap(0,1))\big)$, where $f$ is a one-to-one and measurable function from $(0,1)$ to $\R$ and $\sigma(\bar{F})\subset[0,1]$. In the appendix A we prove that $A\leftrightarrow\Pr A^+$ if and only if $A\leftrightarrow\Pr\bar{A}^+$, having denoted by $A^+$ and $\bar{A}^+$ the Neumark operators $\int\lambda dE^+_{\lambda}$ and $\int\lambda d{\bar E}^+_{\lambda}$ associated to $F$ and $\bar{F}$ respectively.
Therefore, in what follows we restrict ourself, without loss of generality, to POV measures with spectrum in $[0,1]$. 
\renewcommand{\thesection}{\Roman{section}}
 
\subsection{The general case}
In the present subsection we generalize theorem 5 of Ref. 18. In particular in the appendix B we prove the following theorem:

\begin{teo}

Let $F$ be a commutative POV measure with spectrum in $[0,1]$ and $A$ the sharp reconstruction of $F$.  
Let $E^+$ be the extension of $F$ whose existence is asserted by Neumark's theorem, $A^+$ the Neumark operator $\int\lambda dE^+_{\lambda}$. Then, to each bounded and measurable function $f:[0,1]\to\R$ there corresponds a function $G_f:[0,1]\to[0,1]$ such that 
$$B:=P^+f(A^+)_{\vert \cal H}=G_f(A).$$
\end{teo}

\noindent
In Ref. 18, it was analyzed the case $f=\lambda$ and it was given an example (see example 1 below) where the operators $A$ and $B$ coincide up to a bijection, thanks to the injectivity of $G_f$. 
\begin{example}

Let us consider the POV measure
\begin{equation}
F(\Delta)=\begin{cases}
\Phi & \text{if $1\in\Delta$ and $0\not\in\Delta$} \\
C={I}-\Phi & \text{if $1\not\in\Delta$ and $0\in\Delta$}\\
{I} & \text{if ${1,0}\in\Delta$}\\
0 & \text{if $1\not\in\Delta$ and $0\not\in\Delta$} .
\end{cases}
\end{equation} 
where $\Phi$ is a bounded self-adjoint operator such that $0\leq\Phi\leq I$.
We can easily find a family of probability measures $\omega_{(\cdot)}(\lambda):{\cal B}([0,1])\to[0,1]$, $\lambda\in[0,1]$, such that $\omega_{\Delta}(\Phi)=F(\Delta)$. It is sufficient to choose
\begin{equation}
\omega_{\Delta}(\lambda)=\begin{cases}
\lambda & \text{if $1\in\Delta$ and $0\not\in\Delta$} \\
1-\lambda & \text{if $1\not\in\Delta$ and $0\in\Delta$}\\
1 & \text{if ${1,0}\in\Delta$}\\
0 & \text{if $1\not\in\Delta$ and $0\not\in\Delta$} .
\end{cases}
\end{equation}
Therefore the triplet $(F,\Phi,\omega_{(\cdot)}(\lambda))$ satisfies the thesis of von Neumann theorem.
%The probability distribution associated to the probability measure $\omega_{(\cdot)}(\lambda)$ is
%\begin{equation}
%\omega_t(\lambda)=\begin{cases}
%0 & t\leq 0\\
%1-\lambda & 0<t\leq 1\\
%1 & t> 1
%\end{cases}
%\end{equation}
Now we show that $\Phi$ coincides with the projection $B:=P^+A^+_{\vert\cal H}=\int tdF_t$ of the Neumark operator $A^+$ corresponding to the extension $E^+$ of $F$. Indeed, we can follow the proof of Theorem 4, with the operator $A$ replaced by the operator $\Phi$, and $f=\lambda$. Then, we get $G^{(\phi)}_f(\Phi)=B$, where $G^{(\Phi)}_f(\lambda)=\int_{0}^{1^+}t\,\,d_t[\omega_t(\lambda)]$. Moreover, 
$$G^{(\Phi)}_f(\lambda)=\int_{0}^{1^+}t\,\,d_t[\omega_t^{\Phi}(\lambda)]=\lambda$$
hence,
$$B=\Phi.$$

%\noindent
%One has:
%$$\langle Bx,y\rangle=\int_{0}^{1^+}t\,d\langle F_tx,y\rangle=\langle \Phi x,y\rangle\quad \text{for each}\quad x,y\in \cal H$$
%hence,
%$$B=\Phi.$$

%\noindent
%Now, we want to calculate $G^{(\Phi)}(\lambda)$ in order to show that actually it is $G^{(\Phi)}(\Phi)=B$.

%\noindent
%One has:
%$$G^{(\Phi)}(\lambda)=\int_{0}^{1^+}t\,\,d_t[\omega_t^{\Phi}(\lambda)]=\lambda$$
%which applied to the operator $\Phi$ gives:
%$$G^{(\Phi)}(\Phi)=\Phi=B.$$
\noindent
Now, let us consider the sharp reconstruction $A$ corresponding to $F$. By applying theorem 4 with $f=\lambda$, we get 
$$G_f(A)=B=\Phi.$$
Since the triple $(F,\Phi,\omega_{(\cdot)}(\lambda))$ satisfies the thesis of von Neumann theorem, it follows (by theorem 2) that there exists a function $g:\sigma(\Phi)\to[0,1]$ such that $g(\Phi)=A$. In Ref. 16 it was shown that $g$ is injective. Therefore, $g(\Phi)=g[G_f(A)]=A$ and $g[G_f(\lambda)]=\lambda$ a.e. with respect to the PV measure $E^A$ corresponding to $A$. This means that $G_f(\lambda)$ is injective in $\sigma(A)$ so that the sharp reconstruction $A$ of $F$ and the projection $\Phi$ of the Neumark operator $A^+$ are equivalent, hence $A\leftrightarrow Pr A^+$. 
\end{example}
\noindent
Next we recall an example, used by Grabowski$^{23}$ to analyze the concept of unsharp observable, which we use as an example of a POV measure whose sharp reconstruction coincides (up to a bijection) with the projection of the Neumark operator $A^+$ (hence $A\leftrightarrow \Pr A^+$).
\begin{example}
Let ${\cal H}={\C}^3$ be the Hilbert space for a system with spin $J=1$, and $E_{-1},\,E_{0},\,E_{1}$ the projections corresponding to the eigenvectors of the spin observable $J_3=\sum_{m=-J}^{J}mE_{m}$.
Let us consider the position operator $Q:L^2(\R)\to L^2(\R)$ of a particle in $\R$, the corresponding PV measure $E(\cdot):{\cal B}(\R)\to{\cal E}(L^2(\R))$ and a vector $\psi\in L^2(\R)$ such that $\langle\psi,Q\psi\rangle=0$. Let us define the commutative POV measure  
$$F(\Delta)=\sum_{m=-1}^1\langle\psi,E(\Delta+m)\psi\rangle E_m$$
in $C^3$.

\noindent
The projection of the Neumark operator coincides with $J_3$. Indeed, 
$$\int_{\R}\lambda dF_{\lambda}=\sum_m E_m\int_{\R}\lambda\vert\psi(\lambda-m)\vert^2 d\lambda=\sum_m m E_m=J_3.$$
Because of the maximality of $J_3$ (that is, if $J_3=g(A)$ then $g$ is one-to-one) and of theorem 4, the sharp reconstruction $A$ must be equivalent to $J_3$, $A\leftrightarrow J_3$. 
\end{example} 

\subsection{The finite dimensional case}
In this section we show that, in the finite dimensional case, the sharp reconstruction of a given commutative POV measure is equivalent to the projection of the Neumark operator in the sense specified by definition 9.    

\noindent
Therefore in what follows we restrict ourself to the finite dimensional case, and consider an $n$-dimensional Hilbert space $\cal H$. Definition 1 becomes:
\noindent
\begin{defin}

For a finite or countable outcome set $K\subset\R$, a POV measure $F$ is an application $F:K\to\{F_k\}_{k\in K}$, also denoted by $\{F_k\}_{k\in K}$, where $\{F_k\}_{k\in K}$ is a set of positive self-adjoint operators acting on a finite dimensional Hilbert space $\cal H$, such that:
\begin{equation*}
\sum_{k\in K}F_k=\bf{1}.
\end{equation*}
\end{defin}

\noindent
In Ref. 34, it is given a procedure for obtaining the sharp reconstruction $A$ corresponding to a commutative POV measure $F$ with a finite outcome set. Being interested in the spectral measure corresponding to $A$, we outline the procedure for its construction which we generalize to the case of an infinite but countable outcome set.

\noindent
Let us consider a commutative POV measure $F:\{k_1,\dots,k_m,\dots\}\to\{F_{k_1},\dots,F_{k_m},\dots\}$ in an $n$-dimensional Hilbert space $\cal H$.
Let $E^{(i)}_j$, $j=1,2,\dots,n$, be a set of $n$ one-dimensional projections corresponding to a base for $\cal H$ which diagonalizes the operator $F_{k_i}$, and let $\lambda^{(i)}_j$ be the corresponding eigenvalues. %the spectral measure and the corresponding eigenvalues associated to the self-adjoint operator $F_{k_i}$. 
 Notice that the eigenvalues are not necessarily distinct. Moreover, because of the commutativity relations:
\begin{equation*}
[F_{k_i},F_{k_j}]={\bf{0}},\quad i,j=1,\dots,m,\dots
\end{equation*}
we can assume 
\begin{equation*}
E_j:=E^{(i)}_j=E^{(l)}_j,\quad i,l=1,\dots,m,\dots,\quad j=1,\dots,n
\end{equation*}
so that 
\begin{equation}
F_{k_i}=\sum_{j=1}^n{\lambda^{(i)}_j}E_j.
\end{equation}
Here, the real number $\lambda^{(i)}_j$ is the eigenvalue of $F_{k_i}$ which corresponds to the projection $E_j$.

\noindent
Next, for any $j\in\{1,\dots,n\}$, let us consider the sequence $\{\lambda^{(1)}_j,\dots,$ $\lambda^{(m)}_j,\dots\}$. There exists a set of projection operators $E^A_1,E^A_2,\dots,E^A_N$, where $N\leq n$, such that the sequences $\{\lambda^{(i)}_j\}_{i=1,\dots,m,\dots}$, $j=\{1,\dots,N\}$, corresponding to the projections $E^A_j$, $j=1,\dots,N$, are distinct, i.e., for every couple of indexes $(j,l)$ there exists an index $i$ such that $\lambda_j^{(i)}\neq\lambda_l^{(i)}$. Indeed, if the sequences $\{\lambda^{(i)}_j\}_{i=1,\dots,m,\dots}$ and $\{\lambda^{(i)}_l\}_{i=1,\dots,m\dots}$, corresponding to the projections $E_j$ and $E_l$, are equal, we can replace in (11) $E_j$ with the projection $E_j+E_l$ and skip the $l$-th term. Iterating this procedure, after relabeling the indexes, we get $F_{k_i}=\sum_{j=1}^N\lambda_j^{(i)}E^A_j$ for every $i\in\N$, for an integer $N\le n$. The resulting sequences $\{\lambda^{(i)}_j\}_{i=1,\dots,m,\dots}$, $j=\{1,\dots,N\}$, corresponding to the projections $E^A_j$, $j=1,\dots,N$, are distinct.

\noindent
Moreover, if $n(i)$ is the number of distinct eigenvalues of $F_{k_i}$, we have $n(i)\leq N$ for every $i\in\N$. 

\noindent
The sharp reconstruction $A$ of $F$ is defined$^{34}$ (up to bijections) as follows
\begin{equation*}
A=\lambda_1E^A_1+\lambda_2E^A_2+\cdots+\lambda_NE^A_N,
\end{equation*}
where $\{\lambda_i\}_{i=1,\dots,N}=\sigma(A).$
The POV measure $F$ can be interpreted as a randomization of $A$. Indeed, the functions 
\begin{align*}
f^A_{k_i}:\{\lambda_1,\lambda_2,\dots,\lambda_N\} & \to\{\lambda^{(i)}_1,\lambda^{(i)}_2,\dots,\lambda^{(i)}_N\}\in\R\\
\lambda_j & \mapsto\lambda^{(i)}_j 
\end{align*}

%continuare
\noindent
are such that
\begin{description}
\item {i)} $f^A_{k_i}(A)=F_{k_i}$;
\item {ii)} $\sum_{i=1}^{\infty}f^A_{k_i}(\lambda)=1$\quad\text{for every $\lambda\in\sigma(A)$.}
\end{description}
Item i) is quite obvious while item ii) comes from definition 10. Notice that the functions $f^A_{k_i}$ are not generally one-to-one.

\noindent
The following theorem, which corresponds to the first part of theorem 2, summarizes what said above.
\begin{teo}
A POV measure $F:\{k_1,\dots,k_m,\dots\}\to\{F_{k_1},\dots,F_{k_m},\dots\}$ is commutative if and only if there exist a PV measure $E^A:\{\lambda_1,\dots,\lambda_N\}\to\{E^A_{1},\dots,E^A_{N}\}$ and a set of functions $f^A_{k_i}:\{\lambda_1,\dots,\lambda_N\}\to\{\lambda^{(i)}_1,\dots,\lambda^{(i)}_N\}$, $i=1,\dots,m,\dots$, such that
\begin{description}
\item {i)} $f^A_{k_i}(A)=F_{k_i}$;
\item {ii)} $\sum_{i=1}^{\infty} f^A_{k_i}(\lambda)=1$\quad\text{for every $\lambda\in\{\lambda_1,\dots,\lambda_N\}$};
\item {iii)} for every couple $(\lambda_i,\lambda_j)$ there exists an index $l\in\N$ such that $f^A_{k_l}(\lambda_i)\neq f^A_{k_l}(\lambda_j)$,
\end{description}
where $A=\sum_{j=1}^N\lambda_jE^A_j$ is the self-adjoint operator corresponding to $E^A$.
\end{teo}

\noindent
Following H. Martens and W. M. de Muynck$^{20,21}$ (see also definition 11), we summarize the relationship between POV measures and sharp reconstruction, expressed by items i) and ii) in theorem 5, by writing
$$E^A\to F.$$
In the finite dimensional case, the second part of theorem 2 becomes:
\begin{teo}
If $B$ and $\{f^B_{k_i}\}_{i=1,\dots,m,\dots}$ are respectively a self-adjoint operator and a family of functions $f^B_{k_i}:\sigma(B)\to\{\lambda^{(i)}_1,\dots,\lambda^{(i)}_N\}$, such that
\begin{description}
\item {i)} $f^B_{k_i}(B)=F_{k_i}$;
\item {ii)} $\sum_{i=1}^{\infty} f^B_{k_i}(\lambda)=1$\quad\text{for every $\lambda\in\sigma(B)$};
\end{description}
then, there exists a function $g$ such that 
$$g(B)=A$$
where, $A$ is the sharp reconstruction of $F$.
\end{teo}

\noindent
Now we can proceed to prove the main result of the paper, theorem 7. By Neumark's theorem, the POV measure $\{F_{k_j}\}_{j=1,\dots,m,\dots}$ can be extended to a PV measure $\{E_j^+\}_{j=1,\dots,m,\dots}$ in a Hilbert space ${\cal H}^+\supset\cal H$ such that ${P^+E_j^+}_{\vert\cal H}=F_{k_j}$. We show that there exists a one-to-one function $f$ such that the projection $B=P^+B^+_{\vert\cal H}=\sum_jf(k_j)F_{k_j}$ of the Neumark operator $B^+=f(A^+)=\sum_jf(k_j)E^+_j$ is equivalent to $A$, i.e., there exists a one-to-one function $G_f$ such that $B=G_f(A)$. 

\noindent
The following lemma, which we prove in appendix C, is the key of the proof of theorem 7.  

\begin{lem}[see appendix C]

A matrix of real numbers:
\begin{equation}\begin{pmatrix}
\lambda_1^{(1)} & \lambda_1^{(2)} & \dots & \lambda_1^{(m)} & \dots\\
\lambda_2^{(1)} & \lambda_2^{(2)} & \dots & \lambda_2^{(m)} & \dots\\
\hdotsfor{5}\\
\lambda_N^{(1)} & \lambda_N^{(2)} & \dots & \lambda_N^{(m)} &\dots
\end{pmatrix}\end{equation}
such that:
\begin{description}
\item{i)} for every couple of indexes $(i,j)$ there exists an index $l\in\N$ such that $\lambda_i^{(l)}\neq\lambda_j^{(l)}$;
\item{ii)} $\sum_{i=1}^{\infty}\lambda_j^{(i)}=1$.
\end{description}
defines a compact operator $T:l_{\infty}\to{\C}^N$ with the property that there exists a real vector $\{k_1,k_2,\dots,k_m,\dots;\,\,k_i\neq k_j,\,\,i\neq j\}\in l_{\infty}$ such that the elements of the image vector 
\begin{equation}
\begin{pmatrix}
a_1\\
a_2\\
\vdots\\
a_N
\end{pmatrix}:=
\begin{pmatrix}
\lambda_1^{(1)} & \lambda_1^{(2)} & \dots & \lambda_1^{(m)} & \dots\\
\lambda_2^{(1)} & \lambda_2^{(2)} & \dots & \lambda_2^{(m)} & \dots\\
\hdotsfor{5}\\
\lambda_N^{(1)} & \lambda_N^{(2)} & \dots & \lambda_N^{(m)} & \dots
\end{pmatrix}
\begin{pmatrix}
k_1\\
k_2\\
\vdots\\
k_m\\ 
\vdots
\end{pmatrix},
\end{equation}
are distinct real numbers, i.e., $a_i\neq a_j$ if $i\neq j$. 
\end{lem}
\begin{teo}
Let $F:\{k_1,\dots,k_m,\dots\}\to\{F_{k_1},\dots,F_{k_m},\dots\}$ be a commutative POV measure with spectrum in $[0,1]$, $A=\lambda_1E^A_1+\dots+\lambda_NE^A_N$ its sharp reconstruction, $E^+:\{k_1,\dots,k_m,\dots\}\to\{E^+_{k_1},\dots, E^+_{k_m},\dots\}$ an extension of $F$ whose existence is asserted by Neumark's theorem and $A^+$ the Neumark operator $\sum_{j=1}^{\infty} k_jE^+_{k_j}$. Then $A\leftrightarrow\Pr A^+$. %there exists a bounded operator $B^+=\sum_j^{\infty}\lambda_jE^+_{k_j}$ whose projection $P^+B^+_{\vert\cal H}$ is equivalent to the sharp reconstruction $A$. That is, there exists a bijective function $g$ such that:
%\begin{equation*}
%g(A)=P^+B^+_{\vert\cal H}
%\end{equation*}
\end{teo}
\begin{proof}
Let us consider a bounded function $f:\{k_1,\dots,k_m,\dots\}\to \R$ and the bounded operator $B^+=f(A^+)=\sum_j^{\infty}f(k_j)E^+_{k_j}$.
By lemma 1, we get 
\begin{align}
P^+B^+_{\vert\cal H}&=P^+\sum_j^{\infty}f(k_j) {E^+_{k_j}}_{\vert\cal H}=\sum_j^{\infty}f(k_j)F_{k_j}=\sum_{j=1}^{\infty}f(k_j)\sum_{i=1}^N\lambda_i^{(j)}E^A_i\notag\\
&=\sum_{i=1}^N\bigg(\sum_{j=1}^{\infty}\lambda_i^{(j)}f(k_j)\bigg)E^A_i=\sum_{i=1}^NG_f(\lambda_i)E^A_i=G_f(A)
\end{align} 
where $A$ is the sharp reconstruction of $F$ and  
\begin{equation*}
G_f(\lambda_i):=\sum_{j=1}^{\infty}\lambda_i^{(j)}f(k_j)\leq\sup_j{\vert f(k_j)\vert},
\quad \lambda_i\in\sigma(A), \quad i=1,\dots,N.
\end{equation*}
In matrix form, we can write
\begin{equation}
\begin{pmatrix}
G_f(\lambda_1)\\
G_f(\lambda_2)\\
\vdots\\
\vdots\\
G_f(\lambda_N)
\end{pmatrix}=
\begin{pmatrix}
\lambda_1^{(1)} & \lambda_1^{(2)} & \dots & \lambda_1^{(m)} & \dots\\
\lambda_2^{(1)} & \lambda_2^{(2)} & \dots & \lambda_2^{(m)} & \dots\\
\hdotsfor{5}\\
\hdotsfor{5}\\
\lambda_{N}^{(1)} & \lambda_{N}^{(2)} & \dots & \lambda_{N}^{(m)} & \dots
\end{pmatrix}
\begin{pmatrix}
f(k_1)\\
f(k_2)\\
\vdots\\
f(k_m)\\
\vdots 
\end{pmatrix}
\end{equation}
Moreover, by theorem 5 (items ii and iii), we have:
\begin{description}
\item{i)} for every couple of indexes $(i,j)$ there exists an index $l\in\N$ such that $\lambda_i^{(l)}\neq\lambda_j^{(l)}$;
\item{ii)} $\sum_{j=1}^{\infty}\lambda_i^{(j)}=1$\quad\text{for every}\quad $i\in\{1,2,\dots,m\}$. 
\end{description}
By lemma 1, there exists a vector $\{\lambda^+_1,\dots,\lambda^+_m,\dots;\lambda^+_i\neq\lambda^+_j,i\neq j\}\in l_{\infty}$ such that the function $f$ defined by $f(k_1)=\lambda^+_1,\dots,f(k_m)=\lambda^+_m,\dots$ and the function $G_f$ are one-to-one, i.e., $f(k_i)\neq f(k_j)$ and $G_f(\lambda_i)\neq G_f(\lambda_j)$ if $i\neq j$. By equation (14), $P^+f(A^+)=P^+B^+_{\vert\cal H}=G_f(A)$ and then $A\leftrightarrow \Pr A^+$.
\end{proof}
\noindent
\begin{example}

Let ${\cal H}={\C}^2$ be the Hilbert space for a system with spin $J=1/2$. Let $P_{+},\,P_{-}$ be the projections corresponding to the eigenvectors of the spin observable $J_z=1/2P_{+}-1/2P_{-}$.
Let us consider the commutative POV measure $F:\{1/2,-1/2\}\to\{F_1=(1-\epsilon) P_++\delta P_-,\,F_2=\epsilon P_++(1-\delta)P_-\}$, $\epsilon+\delta\neq 1$, which can be interpreted$^{\,25,35}$ as the representation of the measurement of the spin in the $z$ direction where a `spin-up' is registered as `spin-down' with probability $\epsilon$ and a `spin-down' is registered as `spin-up' with probability $\delta$. The sharp reconstruction of $F$ is
\begin{equation*}
A=1P_++2P_-.
\end{equation*}
The functions $f_1$ and $f_2$ connecting $F$ and $A$ are defined as follows
\begin{align*}
f_1(1)=1-\epsilon\,\, & ;\quad f_1(2)=\delta\\
f_2(1)=\epsilon\,\, & ;\quad f_2(2)=1-\delta
\end{align*}
By Neumark's theorem there exists an extended Hilbert space ${\cal H}^+$ and an orthogonal resolution of the identity $\{E^+_1,E^+_2\}$ in ${\cal H}^+$ such that ${P^+E^+_i}_{\vert\cal H}=F_i$.
It is easy to see that the Neumark operator $A^+=1/2E^+_1-1/2E^+_2$, corresponding to the POV measure $F$, is such that its projection $P^+A^+_{\vert\cal H}$ coincides, up to a bijection, with the sharp reconstruction $A$.

\noindent
Indeed, 
\begin{align*}
P^+A^+_{\vert\cal H}=1/2P^+E^+_1-1/2P^+E^+_2=1/2F_1-1/2F_2\\
=1/2[(1-\epsilon) P_++\delta P_-]-1/2[\epsilon P_++(1-\delta)P_-]\\
=(1/2-\epsilon) P_++(-1/2+\delta) P_-=f(A)
\end{align*}
where 
\begin{equation*}
f(1)=1/2-\epsilon;\quad f(2)=-1/2+\delta.
\end{equation*}
\end{example}
\begin{example}

Let ${\cal H}={\C}^3$ be the Hilbert space for a system with spin $J=1$. Let $E_{-1},\,E_{0},\,E_{1}$ be the projections corresponding to the eigenvectors of the spin observable $J_3=\sum_{m=-1}^{1}mE_{m}$. Let us consider the POV measure $\{1,2,3\}\to\{F_1=1/2E_{-1}+1/2E_0+1/4E_1,\,\,F_2=1/5E_{-1}+1/5E_0+1/4E_1,\,\,F_3=3/10E_{-1}+3/10E_0+1/2E_1\}$. The corresponding sharp reconstruction is $A=1(E_{-1}+E_0)+2E_1$. The projection of the Neumark operator $A^+$ corresponding to $F$ is $P^+A^+_{\vert\cal H}=1F_1+2F_2+3F_3=9/5(E_{-1}+E_0)+9/4E_1=f(A)$ where $f$ is the one-to-one function such that $f(1)=9/5$ and $f(2)=9/4$. Notice that $A$ is a function of $J_3$ ($A=g(J_3)$ where $g(-1)=1,\,g(0)=1,\,g(1)=2$). 
\end{example}
\noindent
\renewcommand{\thesection}{\Roman{section}.}
\section{Neumark's Theorem and Nonideal Quantum Measurement}
In this section we briefly recall the definition of non-ideal quantum mea-surement$^{20,21}$ and its connection with corollary 1, and analyze some implications of theorems 7 and 6 to this connection (corollary 3). Moreover we comment (definition 13, theorem 9) on the relationships between different definitions of unsharpness. 

\noindent
The concept of non-ideal quantum measurement of a PV measure $E$ by means of a POV measure $F$ is defined as follows:
\begin{defin}[see Ref.s 20,21]
The POV measure $F:K\to\{F_k\}_{k\in K}$ with a finite or countably infinite outcome set $K$ is said to be an unsharp version of the PV measure $E:L\to\{E_l\}_{l\in L}$ if there exists a set of non-negative real numbers $\{\lambda_l^{(k)}\}_{k\in K,\,l\in L}$ such that: 
\begin{description}
\item {i)} $\sum_{k\in K}\lambda_l^{(k)}=1$,
\item {ii)} $F_k=\sum_{l\in L}\lambda_l^{(k)}E_l$
\end{description}
\end{defin}
\begin{rem}
Notice that definition 11 is equivalent to requiring that for each operator $B=\sum_{l\in L}\gamma_lE_l$, with $\{\gamma_l\}_{l\in L}$ set of distinct real numbers, there exists a set of functions $\{f^B_k\}_{k\in K}$ such that $f^B_k(B)=F_k$ (e.g. $f^B_k(\gamma_l)=\lambda_l^{(k)}$). 
\end{rem}
\noindent
In this paper we do not distinguish between a self-adjoint operator $B$ and the corresponding PV measure $E^B$. Therefore, we say that an unsharp observable, represented by a POV measure $F$, is the unsharp version of a sharp observable, represented by a self-adjoint operator $B$, if $F$ is the unsharp version of the PV measure $E^B$ corresponding to $B$. 
In particular, a commutative POV measure $F$ is the unsharp version of its sharp reconstruction $A$. 
\begin{example}
The POV measure $F$ in example 4 is an unsharp version of the spin observable $J_3=\sum_{m=-1}^1mE_i$. Indeed, by setting:
\begin{equation*}
\begin{pmatrix}
\lambda_1^{(1)} & \lambda_1^{(2)} & \lambda_1^{(3)}\\
\lambda_0^{(1)} & \lambda_0^{(2)} & \lambda_0^{(3)}\\
\lambda_{-1}^{(1)} & \lambda_{-1}^{(2)} & \lambda_{-1}^{(3)}
\end{pmatrix}=
\begin{pmatrix}
1/2 & 1/5 & 3/10\\
1/2 & 1/5 & 3/10\\
1/4 & 1/4 & 1/2
\end{pmatrix}
\end{equation*}
we get $E\to F$, where $E=\{E_{-1},E_{0},E_{1}\}$. Moreover, by remark 1, there exists a family of functions $f_i$ such that $f_i(J_3)=F_i$, then, by theorem 6, there is a function $g$ such that $g(J_3)=A$, where $A=E_{-1}+E_0+2E_1$ is the sharp reconstruction of $F$. 
\end{example}
\noindent
Before giving the connection between definition 11 and corollary 1 we state the latter in the finite dimensional case.
\begin{coro}[see Ref.s 3,9,21]
For every POV measure $\{F_k\}_{k\in K}$ on $\cal H$, with a finite or countable outcome set $K$, there is a Hilbert space ${\cal H}'$, a density operator ${\rho}'$ on ${\cal H}'$, and a PV measure $\{E^+_k\}_{k\in K}$ on ${\cal H}\otimes{\cal H}'$ such that
\begin{equation*}
F_k=Tr_{{\cal H}'}({\rho}'E^+_k)
\end{equation*}
\end{coro}

\noindent
The connection mentioned above is summarized by the following theorem:
\begin{teo}[see Ref.s 20,21]
Let $\bar{A}=\sum_aa\bar{E}_a$ be a self-adjoint operator on $\cal H$ and $A^+=\sum_kkE^+_k$ a self-adjoint operator on ${\cal H}\otimes{\cal H}'$. If there exist a self-adjoint operator $T=\sum_l l E'_l$ on ${\cal H}'$, and a function $k(a,l)$ such that $A^+=k(A,T)$, then, for every density operator ${\rho}'$ on ${\cal H}'$,
\begin{equation*}
\bar{E}\to F,
\end{equation*}
where $\bar{E}$ and $F$ are respectively the PV measure corresponding to $\bar{A}$ and the POV measure defined by $F_k=Tr_{{\cal H}'}({\rho}'E^+_k)$.  
\end{teo}
\noindent 
The following corollary is a consequence of theorems 7 and 6.
\begin{coro}
Under the hypothesis of theorem 8, let us consider the density operator ${\rho}'$, the corresponding POV measure $F_k=Tr_{{\cal H}'}({\rho}'E^+_k)$ and its sharp reconstruction $A$. If the function $f(a)=\sum_lk(a,l)Tr_{{\cal H}'}({\rho}'E'_l)$ is one-to-one, then $A\leftrightarrow\bar{A}$. Conversely, if $A\leftrightarrow\bar{A}$ then there exists a one-to-one function $h$ such that the function $r(a)=\sum_lh(k(a,l))Tr_{{\cal H}'}({\rho}'E'_l)$ is one-to-one. 
\end{coro}
\begin{proof}
Assume $f$ is one-to-one. Since $\bar{E}\rightarrow F$ there exists (by theorem 6) a function $g$ such that 
$$g(\bar{A})=A,$$
where $A=\sum\lambda_{n}E_n$ is the sharp reconstruction of $F$. Moreover, proceeding as in equation (14) in theorem 7, we get a function $G$ such that 
\begin{equation}
G(A)=\sum_kkF_k. 
\end{equation}

\noindent
Hence, 
$$G[g(\bar{A})]=\sum_kkF_k=Tr_{{\cal H}'}[{\rho}'A^+]$$
$$=\sum_a\bigg(\sum_l k(a,l)Tr_{{\cal H}'}({\rho}'E'_l)\bigg)\,\bar{E}_a=f(\bar{A})$$
which shows that $f$ is one-to-one if and only if both $g$ and $G$ are one-to-one. The thesis comes from the fact that $f$ is one-to-one. Conversely, if $A\leftrightarrow \bar{A}$ then there is a one-to-one function $g$ such that $A=g(\bar{A})$. Furthermore, by theorem 7, there exist two one-to-one functions $h$ and $G_h$ such that $$G_h(A)=\sum_kh(k)F_k=Tr_{{\cal H}'}[{\rho}'h(A^+)]=r(\bar{A})$$
where, 
$$r(a)=\sum_lh(k(a,l))Tr_{{\cal H}'}({\rho}'E'_l)$$
which means that $r=g^{-1}\circ G_h $ is one-to-one (the symbol $\circ$ denotes the operation of composition between functions).
\end{proof}
\noindent
We recall that another definition of ``unsharpness'' is the following$^{22-25}$: 
\begin{defin}
The observable represented by the POV measure $F:K\to\{F_k\}_{k\in K}$ is an unsharp version of the observable represented by a self-adjoint operator $B$, if there exists a sequence of real numbers $\{\gamma_k\}_{k\in K}$ such that 
$$B=\sum_k\gamma_k F_k.$$
\end{defin}

\begin{example}
From example 3 we have that the observables $P_{+}$, $P_{-}$ and $f(A)$ can be written as $P_{+}=\frac{1-\delta}{1-\epsilon-\delta}F_1+\frac{-\delta}{1-\epsilon-\delta}F_2$, $P_{-}=\frac{-\epsilon}{1-\epsilon-\delta}F_1+\frac{1-\epsilon}{1-\epsilon-\delta}F_2$, $f(A)=1/2F_1-1/2F_2$. Therefore, $F$ is an unsharp version of $P_{+}$, $P_{-}$ and $f(A)$. Moreover, this shows that all the observables which are function of the sharp reconstruction $A$ can be represented as a sum of the kind $\sum \gamma_iF_i$. Notice that $F$ is not an unsharp version of $P_{+}$ and $P_{-}$ in the sense of definition 11. 
\end{example}
\noindent
Grabowski$^{23}$ conjectured that definition 12 is equivalent to definition 11 but Uffink$^{25}$ observed that this is false since, according to definition 11, any unsharp version of a sharp observable $B$ must be commutative and this is not true for definition 12. Now we show that also in the case we restrict ourself to commutative POV measures Grabowski's conjecture is false.
\noindent  
Indeed, let us consider the POV measure $F$ in example 4. From example 5 it follows that $A=g(J_3)$ where $A$ is the sharp reconstruction of $F$ and $g$ is not one-to-one since $g(-1)=g(0)=1$. Assume $F$ unsharp version of $J_3$ in the sense of definition 12, i.e., 
$$J_3=\sum_{i=1}^3\gamma_i F_{i}.$$
By proceeding as in equation (14), we get a function $G(j)=\sum_{i}\lambda_j^{(i)}\gamma_i$ such that
$$G(A)=\sum_{i=1}^3\gamma_i F_{i}.$$
Then,   
$$G(A)=J_3 \quad\text{and}\quad g(J_3)=A$$
hence,
$$G(g(J_3))=J_3$$ 
which means that $g$ is one-to-one and contradicts the hypothesis. Therefore, $F$ is an unsharp version of $J_3$ according to definition 11 but not according to definition 12.

The following definition is a generalization of both definitions 11 and 12.
\begin{defin}
The observable represented by the POV measure $F:K\to\{F_k\}_{k\in K}$ is an unsharp version of the observable represented by the operator $B$ if there exist a function $h$ and a sequence of real numbers $\{\gamma_k\}_{k\in K}$ such that
$$h(B)=\sum_k\gamma_k F_k$$
\end{defin} 
\begin{teo}
If $F$ is an unsharp version of $B$ according to definitions 11 or 12 then, it is an unsharp version of $B$ according to definition 13.
\end{teo}
\begin{proof}
Let $E^B$ be the PV measure corresponding to the self-adjoint operator $B$. If $E^B\to F$ then there exists a function $g$ such that $g(B)=A$, where $A$ is the sharp reconstruction of $F$. Moreover, proceeding as in equation (14), we get, for any bounded function $f:K\to\R$, a function $G_f$ such that $G_f(A)=\sum_k f(k)F_k$, hence
$$h(B):=G_f(g(B))=\sum_k f(k)F_k=\sum_k\gamma_k F_k,$$
where we have set $\gamma_k:=f(k)$.
Clearly, if $F$ is an unsharp version of $B$ according to definition 12 then, it is an unsharp version of $B$ according to definition 13 (it is sufficient to choose $h(\lambda)=\lambda$).  
\end{proof}

\noindent
In order to outline the  relationships between definition 11 and definition 12, we introduce the sets ${\cal A}_1$, ${\cal A}_2$, and ${\cal A}'_2$: 
\begin{description}
\item {i)} By definition 11 (see theorem 6), the set of sharp observables of which $F$ is an unsharp version is the set $${\cal A}_1=\{B\in{\cal L}_s{\cal(H)}\,\vert\,\text{there exists a function}\, h\,\, \text{such that}\,\, h(B)=A\},$$ 
\item {ii)} By definition 12 (see theorem 7), the set ${\cal A}_2$ of sharp observables of which $F$ is an unsharp version is a subset of $${\cal A}'_2=\{B\in{\cal L}_s{\cal(H)}\,\vert\,\text{there exists a function}\, g\,\, \text{such that}\,\, B=g(A)\},$$ 
\end {description}
where $A$ is the sharp reconstruction of $F$.

\noindent 
The two sets ${\cal A}_1$ and ${\cal A}'_2$ have a non-empty intersection; e.g., each self-adjoint operator $B$ such that $A=g(B)$, with $g$ one-to-one, belongs to ${\cal A}_1\cap{\cal A}'_2$. Moreover, we have proved that ${\cal A}_1\cap{\cal A}_2\neq\emptyset$ and in particular that the sharp reconstruction of $F$ is contained in ${\cal A}_1\cap{\cal A}_2$.
Now it is clear why definition 13 is a generalization of both definitions 11 and 12; it enlarges ${\cal A}_2$ to a set ${\cal A}$ such that ${\cal A}_1\cup{\cal A}_2\subseteq{\cal A}\subseteq{\cal A}_1\cup{\cal A}'_2$ (notice that, for the POV measure in example 3, ${\cal A}_2={\cal A}'_2$ and then ${\cal A}={\cal A}_1\cup{\cal A}'_2$). A problem to be faced in future investigations is to search for a common meaning of the concepts of unsharpness given by definitions 11 and 12. However, it is worth noticing that the observables in ${\cal A}_2$ can be recovered from $F$ (by appropriately choosing the coefficients in the sum $\sum\gamma_kF_k$) while for the observables in ${\cal A}_1$ this is true only for those observables which are equivalent to the sharp reconstruction $A$ of $F$. Conversely, $F$ can be recovered by each observable $B\in{\cal A}_1$ since to each $B\in{\cal A}_1$ there corresponds a set of functions $f^B_k$ such that $f^B_k(B)=F_k$. Then, we can say that in ${\cal A}_2$ there are observables which contain less information than $F$ and observables which contain the same information as $F$, while, in ${\cal A}_1$ there are observables which contain more information than $F$ and observables which contain the same information as $F$.

\vfill
\section*{Appendices}
\appendix
\section{POV measures with spectrum in $[0,1]$}
\renewcommand{\theequation}{A\arabic{equation}}
\setcounter{equation}{0}
Let $\bar{F}:{\cal B}(\R)\to\cal F(H)$, $\bar{F}(\Delta)=F[f(\Delta\cap(0,1))]$, be the POV measure with spectrum in $[0,1]$ corresponding to the POV measure $F:{\cal B}(\R)\to\cal F(H)$ as stated in section III. If the PV measure $E^+:{\cal B}(\R)\to\cal E(H)$ is the extension of $F$, whose existence is asserted by Neumark's theorem then, $\bar{E}^+(\Delta)=E^+[f(\Delta\cap(0,1))]$ is the extension of $\bar{F}$. Indeed, $P^+\bar{E}^+(\Delta)_{\vert\cal H}=P^+E^+[f(\Delta\cap(0,1))]_{\vert\cal H}=F[f(\Delta\cap(0,1))]=\bar{F}(\Delta)$.
\begin{teo}
Let $F$ and $\bar{F}$ be two commutative POV measures such that $\bar{F}(\Delta)=F[f(\Delta\cap(0,1))]$, let $E^+$ and $\bar{E}^+$ be the corresponding extensions as stated above. Then, $A\leftrightarrow \Pr A^+$ holds if and only if $A\leftrightarrow \Pr \bar{A}^+$ holds.
\end{teo}
\begin{proof}
By the change of measure principle$^{28}$, we get
\begin{equation*}
\begin{split}
\bar{A}^+ & =\int_{-\infty}^{\infty}\lambda d\bar{E}^+_{\lambda}=\int_{-\infty}^{\infty}\lambda \bar{E}^+([\lambda-d\lambda,\lambda))\\
& =\int_{-\infty}^{\infty}\lambda\, E^+[f([\lambda-d\lambda,\lambda))\cap(0,1))]=\int_{(0,1)}\lambda\, E^+[f([\lambda-d\lambda,\lambda))]\\
& =\int_{-\infty}^{\infty}f^{-1}(\lambda)\,dE^+_{\lambda}=f^{-1}(A^+)
\end{split}
\end{equation*}
If $A\leftrightarrow \Pr A^+$ then, there exist two one-to-one, bounded measurable functions $g(\lambda)$ and $h(\lambda)$ such that $g(A)=P^+h(A^+)=P^+h(f(\bar{A}^+))$. Therefore, there exists a one-to-one, bounded, measurable function $H(\lambda)=h(f(\lambda))$ such that $g(A)=P^+H(\bar{A}^+)$ which proves that $A\leftrightarrow \Pr\bar{A}^+$. Conversely, if $A\leftrightarrow \Pr\bar{A}^+$ then there exist two one-to-one, bounded, measurable functions $g(\lambda)$ and $H(\lambda)$ such that $g(A)=P^+H(\bar{A}^+)=P^+H(f^{-1}(A^+))$. Therefore, there exist two one-to-one, bounded, measurable functions $g(\lambda)$ and $h(\lambda)=H(f^{-1}(\lambda))$ such that $g(A)=P^+h(A^+)$ which proves that $A\leftrightarrow \Pr A^+$.
\end{proof}
\section{Proof of theorem 4}
\renewcommand{\theequation}{B\arabic{equation}}
\setcounter{equation}{0}
\begin{proof}
The function $f:[0,1]\to\R$ (we denote by $m$ and $M$ respectively the infimum and the supremum of $f$ in $[0,1]$) and the POV measure $F$ uniquely define$^{27}$ a self-adjoint bounded operator by means of the relation:
\begin{equation}
\langle Bx,x\rangle=\int_{[0,1]}f(t)\,d_t\langle F_tx,x\rangle
\end{equation}
One has:
\begin{align}
\langle Bx,x\rangle &=\int_{[0,1]}f(t)\,d_t\langle F_tx,x\rangle =\int_{[0,1]}f(t)\,d_t\bigg[\int_{[0,1]}\mu^A_t(\lambda)d_{\lambda}\langle E^A_{\lambda}x,x\rangle\bigg]\notag\\
&=\int_{[0,1]}\bigg[\int_{[0,1]}f(t)\,d_t[\mu^A_t(\lambda)]\,\bigg]d_{\lambda}\langle E^A_{\lambda}x,x\rangle\\
&=\langle G_f(A)x,x\rangle \quad\text{for every $x\in\cal H$}\notag
\end{align}
where
\begin{equation}
G_f(\lambda)=\int_{[0,1]}f(t)\,d_t[\mu^A_t(\lambda)],
\end{equation}
$E^A$ and $\mu^A_{(\cdot)}(\lambda)$ are respectively the PV measure corresponding to the sharp reconstruction $A$ and the probability measure whose existence is asserted by theorem 2 and we have denoted by $d_t\int\mu^A_t(\lambda)d_{\lambda}\langle E^A_{\lambda}x,x\rangle$ the integration with respect to the measure $\omega(\cdot)=\int\mu^A_{(\cdot)}(\lambda)d_{\lambda}\langle E^A_{\lambda}x,x\rangle$.

In order to justify the change in the order of integration in equation (B2) we proceed as follows. First we notice that 
$$\omega(\cdot)=\int_{[0,1]}\mu^A_{(\cdot)}(\lambda)d_{\lambda}\langle E^A_{\lambda}x,x\rangle=\langle F(\cdot)x,x\rangle$$  
is, for every $x\in\cal H$, a Lebesgue-Stieltjes measure. %Indeed, by theorem 11 in Ref. 31,   
%\begin{equation}
%\sum_{i=1}^{\infty}\int_{[0,1]}\omega_{(\Delta_i)}(\lambda)\langle E_{\lambda}x,x\rangle=\int_{[0,1]}\sum_{i=1}^{\infty}\omega_{(\Delta_i)}(\lambda)\langle E_{\lambda}x,x\rangle
%\end{equation}
%$\omega_{(\cdot)}(\lambda)$ being a probability measure, thus 
%\begin{description}
%\item {i)} $\mu(\Delta)\geq 0$,\quad for every Borel set $\Delta\in{\cal B}([0,1])$;
%\item {ii)} $\sum_{i=1}^{\infty}\mu(\Delta_i)=\mu(\Delta)$,\quad$\Delta=\cup\Delta_i$,\quad$\Delta_i\cap\Delta_j=\emptyset$, $i\neq j$;
%\item {iii)} $\mu_{[0,1]}=1$
%\end{description}
Therefore, by the definition of Lebesgue-Stieltjes integral$^{36}$,
\begin{align}
\int_{[0,1]} f(t) d_t\omega(t) & =\lim_{\substack{n\to\infty\\\vert\delta_n\vert\to 0}}\sum_{k=1}^{n}f^{(n)}_{k-1}\,\omega\bigg\{t\in[0,1]\,:\,f(t)\in(f^{(n)}_{k-1},f^{(n)}_{k}\big]\bigg\}\notag\\
& =\lim_{\substack{n\to\infty\\\vert\delta_n\vert\to 0}}\sum_{k=1}^{n}f^{(n)}_{k-1}\int_{[0,1]}\mu^A_{(E^{(n)}_{k-1})}(\lambda)\langle E_{\lambda}x,x\rangle\notag\\
& =\lim_{\substack{n\to\infty\\\vert\delta_n\vert\to 0}}\int_{[0,1]}\sum_{k=1}^{n}f^{(n)}_{k-1}\,\mu^A_{(E^{(n)}_{k-1})}(\lambda)\langle E_{\lambda}x,x\rangle
\end{align} 
where it was introduced a sequence of subdivisions $\delta_n=\bigg\{[f_0,f^{(n)}_1],(f^{(n)}_1,f^{(n)}_2],$ $\dots,(f^{(n)}_{n-1},f_n]\bigg\}$, $m=f_0<f_1<\dots<f_n=M$,  of the interval $[m,M]$, such that $\vert\delta\vert=\max_{1\leq k\leq n}\left\{(f^{(n)}_k-f^{(n)}_{k-1})\right\}\to 0$ when $n\to\infty$ and it was set $E^{(n)}_{k-1}=\bigg\{t\in[0,1]\,: f(t)\in(f^{(n)}_{k-1},f^{(n)}_{k}]\bigg\}$.

\noindent
Now let us consider the sequence of functions $$H_n(\lambda)=\sum_{k=1}^{n}f^{(n)}_{k-1}\,\mu^A_{(E^{(n)}_{k-1})}(\lambda).$$
One has 
$$H_n(\lambda)\leq \sup\{\vert f\vert\}\mu_{([0,1])}\,(\lambda)=M<\infty$$
for each $\lambda\in[0,1]$ and $n\in\N$.

\noindent 
Moreover, by the integrability of $f$ with respect to $\mu^A_{(\cdot)}(\lambda)$,
$$\lim_{n\to\infty}H_n(\lambda)=\int_{[0,1]} f(t)\,d_t\mu^A_t(\lambda)=G_f(\lambda).$$
By theorem 11 in Ref. 27,
\begin{align*}
\lim_{n\to\infty}\int_{[0,1]}\sum_{k=1}^{n}f^{(n)}_{k-1}\,\mu_{(E^{(n)}_{k-1})}(\lambda)\langle E_{\lambda}x,x\rangle & =\lim_{n\to\infty}\int_{[0,1]} H_n(\lambda)\langle E_{\lambda}x,x\rangle\\
=\int_{[0,1]}\lim_{n\to\infty}H_n(\lambda)\langle E_{\lambda}x,x\rangle & =\int_{[0,1]}\bigg[\int_{[0,1]}f(t)d_t\mu^A_t(\lambda)\bigg]\langle E^A_{\lambda}x,x\rangle\\
=\int_{[0,1]}G_f(\lambda)\langle E^A_{\lambda}x,x\rangle & =\langle G_f(A)x,x\rangle.
\end{align*}
\noindent
The polarization identity completes the proof.
\end{proof}
\section{Proof of lemma 1}
\renewcommand{\theequation}{C\arabic{equation}}
\setcounter{equation}{0}
\begin{proof}
By item ii), 
$$a_i=\sum_{j=1}^{\infty}\lambda_i^{(j)}k_j\leq\sup_j{\vert k_j\vert}\sum_j^{\infty}\lambda_i^{(j)}= \sup_j{\vert k_j\vert}<\infty$$ 
which means that $T$ is defined everywhere on $l_{\infty}$ and bounded. The compactness of $T$ derives from (see Ref. 29, p. 58) 
$$\sum_{i,j}\vert\lambda_i^{(j)}\vert^2=N<\infty.$$
\noindent
Now, we proceed by induction on $N$.

\noindent
{\bf{Step 1}} (The thesis is true for $N=2$).

\noindent
If $N=2$, (13) becomes
\begin{equation}\begin{pmatrix}
\lambda_1^{(1)} & \lambda_1^{(2)} & \dots & \lambda_1^{(m)} & \dots\\
\lambda_2^{(1)} & \lambda_2^{(2)} & \dots & \lambda_2^{(m)} & \dots\\
\end{pmatrix}
\begin{pmatrix}
k_1\\
k_2\\
\vdots\\
k_m\\
\vdots
\end{pmatrix}=
\begin{pmatrix}
a_1\\
a_2
\end{pmatrix}\end{equation}
We start from a real vector $\{k_1,k_2,\dots,k_m,\dots;\,\,k_i\neq k_j,\,\,i\neq j\}\in l_{\infty}$. Suppose $a_1=a_2$. By item ii), we can assume, without loss of generality, $\lambda_1^{(1)}\neq\lambda_2^{(1)}$ so that, by replacing $k_1$ with $k'_1\neq k_i$, $i\in\N$, we get
\begin{equation}\begin{pmatrix}
\lambda_1^{(1)} & \lambda_1^{(2)} & \dots & \lambda_1^{(m)} & \dots\\
\lambda_2^{(1)} & \lambda_2^{(2)} & \dots & \lambda_2^{(m)} & \dots\\
\end{pmatrix}
\begin{pmatrix}
k'_1\\
k_2\\
\vdots\\
k_m\\
\vdots
\end{pmatrix}=
\begin{pmatrix}
\widetilde{a}_1\\
\widetilde{a}_2
\end{pmatrix}\end{equation}
where $\widetilde{a}_1\neq\widetilde{a}_1$. Indeed,
$$\widetilde{a}_i-a_i=(k_1-k'_1)\lambda_i^{(1)},\quad i=1,2$$
then,
$$\widetilde{a}_1-\widetilde{a}_2=(a_1-a_2)+(k_1-k'_1)(\lambda_1^{(1)}-\lambda_2^{(1)})=(k_1-k'_1)(\lambda_1^{(1)}-\lambda_2^{(1)})\neq 0.$$

\noindent
{\bf Step 2} (Induction on $N$). Suppose that the thesis is true for $N=n$. The case $N=n+1$ reads
\begin{equation}\begin{pmatrix}
a_1\\
a_2\\
\vdots\\
\vdots\\
a_{n+1}
\end{pmatrix}:=
\begin{pmatrix}
\lambda_1^{(1)} & \lambda_1^{(2)} & \dots & \lambda_1^{(m)} & \dots\\
\lambda_2^{(1)} & \lambda_2^{(2)} & \dots & \lambda_2^{(m)} & \dots\\
\hdotsfor{5}\\
\hdotsfor{5}\\
\lambda_{n+1}^{(1)} & \lambda_{n+1}^{(2)} & \dots & \lambda_{n+1}^{(m)} & \dots
\end{pmatrix}
\begin{pmatrix}
k_1\\
k_2\\
\vdots\\
k_m\\
\vdots 
\end{pmatrix}
\end{equation}
Consider the subsystem
\begin{equation}
\begin{pmatrix}
a_1\\
a_2\\
\vdots\\
\vdots\\
a_n
\end{pmatrix}=
\begin{pmatrix}
\lambda_1^{(1)} & \lambda_1^{(2)} & \dots & \lambda_1^{(m)} & \dots\\
\lambda_2^{(1)} & \lambda_2^{(2)} & \dots & \lambda_2^{(m)} & \dots\\
\hdotsfor{5}\\
\hdotsfor{5}\\
\lambda_n^{(1)} & \lambda_n^{(2)} & \dots & \lambda_n^{(m)} & \dots
\end{pmatrix}
\begin{pmatrix}
k_1\\
k_2\\
\vdots\\
k_m\\
\vdots 
\end{pmatrix}
\end{equation}
By the induction hypothesis, there exists a real vector $\{k_1,\dots,k_m,\dots\vert\,\,k_i\neq k_j,\,\,i\neq j\}\in l_{\infty}$ such that the image vector
satisfies the thesis of the lemma, i.e., $a_i\neq a_j$ if $i\neq j$, $i,j=1,\dots,n$. Let us return to consider system (C3) and suppose, without loss of generality, $a_{n+1}=a_1$. By item i), we can assume, without loss of generality, $\lambda_1^{(1)}\neq\lambda_{n+1}^{(1)}$. By replacing $k_{1}$ with a number $k'_1$ such that
\begin{equation}
k'_1\neq
\begin{cases}
k_j & j\in\N\\
k_1-\frac{(a_j-a_i)}{(\lambda_j^{(1)}-\lambda_i^{(1)})} & if\, \lambda_j^{(1)}\neq \lambda_i^{(1)},\,i,j=1,2,\dots,n+1,
\end{cases}
\end{equation}
we get
\begin{equation}\begin{pmatrix}
\lambda_1^{(1)} & \lambda_1^{(2)} & \dots & \lambda_1^{(m)} & \dots\\
\lambda_2^{(1)} & \lambda_2^{(2)} & \dots & \lambda_2^{(m)} & \dots\\
\hdotsfor{5}\\
\lambda_{n+1}^{(1)} & \lambda_{n+1}^{(2)} & \dots & \lambda_{n+1}^{(m)} & \dots
\end{pmatrix}
\begin{pmatrix}
k'_1\\
k_2\\
\vdots\\
k_m\\
\vdots 
\end{pmatrix}=
\begin{pmatrix}
\widetilde{a}_1\\
\widetilde{a}_2\\
\vdots\\
\widetilde{a}_{n+1}
\end{pmatrix}\end{equation}
where $\widetilde{a}_i\neq\widetilde{a}_j$, $i\neq j$.
\noindent
Indeed,
\begin{equation}
\widetilde{a}_i-a_i=(k'_1-k_1)\lambda_i^{(1)}
\end{equation}
and, by subtracting equation (C7) from 
$$\widetilde{a}_j-a_j=(k'_1-k_1)\lambda_j^{(1)},$$
we get 
\begin{equation}
\widetilde{a}_j-\widetilde{a}_i=(k'_1-k_1)(\lambda_j^{(1)}-\lambda_i^{(1)})+(a_j-a_i).
\end{equation}
\noindent
By imposing $\widetilde{a}_j-\widetilde{a}_i\neq 0$ whenever $\lambda_j^{(1)}\neq\lambda_i^{(1)}$, we get the second of (C5). Moreover, if $\lambda_j^{(1)}=\lambda_i^{(1)}$ (which is false if $i=1$ and $j=n+1$) then $\widetilde{a}_j-\widetilde{a}_i=a_j-a_i\neq 0$ for each choice of $k'_1$. 
\end{proof}
\vfill
\section*{References}
\noindent $^1$\footnotesize{E.B.Davies, J.T. Lewis, Commun. Math. Phys., {\bf 17}, 239 (1970) .}

\noindent $^2$\footnotesize{G. Ludwig, \textsl{Foundations of quantum mechanics I}, Springer-Verlag, New York- Heidelberg-Berlin, (1983).}
 
\noindent $^3$\footnotesize{A. S. Holevo, {\sl Probabilistics and statistical aspects of quantum theory,} North Holland, Amsterdam, 1982.}

\noindent $^4$\footnotesize{A. S. Holevo, {\sl Statistical Structure of Quantum Theory}, LNP, m 67, Springer (2001).}

\noindent $^5$\footnotesize{A. S. Holevo, Rep. Math. Phys. {\bf 22}, 385-407 (1985).}

\noindent $^6$\footnotesize{Y. Yamamoto, H.A. Haus, Rev. Mod. Phys. {\bf 58}, 1001 (1986).}

\noindent $^7$\footnotesize{M. Hillery, R. F. O`Connell, M. O. Scully, E. P. Wigner, Phys. Rep. {\bf 106}, 121 (1984).}

\noindent $^8$\footnotesize{E. Prugovecki, \textsl{Stochastic Quantum Mechanics and Quantum Spacetime}, Reidel, Dordrecht (1984).}
 
\noindent $^9$\footnotesize{C. W. Helstrom, {\sl Quantum Detection and Estimation Theory}, Academic Press, New York (1976).}

\noindent $^{10}$\footnotesize{P. Busch, M. Grabowski, P. Lahti, {\sl Operational quantum physics},
LNP m, vol. 31, Springer-Verlag, Berlin (1995).}

\noindent $^{11}$\footnotesize{M. A. Neumark, \textsl{Spectral functions of a symmetric operator}, Izv. Akad. Nauk SSSR Ser. Mat. {\bf 4}, 277-318 (1940).}

\noindent $^{12}$\footnotesize{A.S. Holevo, Trans. Moscow Math. Soc., {\bf 26}, 133 (1972) .}

\noindent $^{13}$\footnotesize{A.S. Holevo, Second Japan URSS Sympos. Probability Theory, vol. 1, Kyoto, 1972, pp.22-40.}

\noindent $^{14}$\footnotesize{S.T.Ali, Lect. notes math., {\bf 905}, 207 (1982).}

\noindent $^{15}$\footnotesize{G. Cattaneo, G. Nistic\`o, J. Math. Phys., {\bf 41}, 4365 (2000).}

\noindent $^{16}$\footnotesize{R. Beneduci, G. Nistic\'o, J. Math. Phys., {\bf 44}, 5461 (2003).}

\noindent $^{17}$\footnotesize{R. Beneduci, J. Math. Phys., {\bf 47} (2006).}

\noindent $^{18}$\footnotesize{R. Beneduci, Int. J. Geom. Methods Mod. Phys., {\bf 3} (2006) (to appear).}

\noindent $^{19}$\footnotesize{M.Reed, B.Simon, {\sl Methods of modern mathematical physics},
Academic Press, NY (1980).}

\noindent $^{20}$\footnotesize{H. Martens, W. M. de Muynck, Found. Phys., {\bf 20}, 357 (1990).}

\noindent $^{21}$\footnotesize{H. Martens, W. M. de Muynck, Found. Phys., {\bf 20}, 255 (1989).}

\noindent $^{22}$\footnotesize{F. E. Schroeck, Jr., Int. J. Theor. Phys., {\bf 28}, 247 (1989).} 

\noindent $^{23}$\footnotesize{M. Grabowski, Found. Phys., {\bf 19}, 923 (1989).}

\noindent $^{24}$\footnotesize{S.T. Al\'i, H.D. Doebner, J. Math. Phys., {\bf 17}, 1105 (1976).}

\noindent $^{25}$\footnotesize{J. Uffink, Int. J. Theor. Phys., {\bf 33}, 199 (1994).}

\noindent $^{26}$\footnotesize{R. Beals, {\sl Topics in Operator Theory}, The University of Chicago Press (Chicago, 1971).}

\noindent $^{27}$\footnotesize{S. K. Berberian, {\sl Notes on Spectral theory}, Van Nostrand Mathematical Studies (Van Nostrand, New Jersey 1966).}

\noindent $^{28}$\footnotesize{N. Dunford, J. T. Schwartz, {\sl Linear Operators} part II, Interscience Publisher (1963).}

\noindent $^{29}$\footnotesize{N. I. Akhiezer and I. M. Glazman, {\sl Theory of Linear Operators in Hilbert Space}, Friedrik Ungar, New York, (1963).}

\noindent $^{30}$\footnotesize{J. von Neumann, {\sl Mathematical Foundations of Quantum Mechanics}, Princeton University Press, Princeton, (1955).}

\noindent $^{31}$\footnotesize{F. Riesz and B. S. Nagy, {\sl Functional Analysis}, Dover, New York, (1990).}

\noindent $^{32}$\footnotesize{M. Lo\`eve, {\sl Probability Theory I}, 4th edition, Springer-Verlag, (1977).}

\noindent $^{33}$\footnotesize{J. F. C. Kingman, S. J. Taylor, {\sl Introduction to Measure And Probability}, Cambridge at the University Press, (1966).}

\noindent $^{34}$\footnotesize{G. Cattaneo, T. Marsico, G. Nistic\`o, G. Bacciagaluppi, Found. Phys., {\bf 27}, 1323 (1997).}

\noindent $^{35}$\footnotesize{P. Bush, F. E. Schroeck, Found. Phys., {\bf 19}, 807 (1989).}

\noindent $^{36}$\footnotesize{T. H. Hildebrandt, {\sl Theory of Integration}, Academic Press, New York, (1963).}  

\end{document}